\documentclass[%
reprint,
superscriptaddress,
 amsmath,amssymb,
 aps,
]{revtex4-2}

\usepackage{graphicx}
\usepackage{dcolumn}

\usepackage{color}
\usepackage{bm}
\usepackage{amssymb}
\usepackage{amsmath}
\usepackage[dvipsnames]{xcolor}

\newcommand{\g}{\textbf}

\newcommand{\ellv}{\bm{\ell} }

\newcommand{\BB}{\bm{B} }

\newcommand{\YY}{\bm{Y} }

\newcommand{\bb}{\bm{b} }

\newcommand{\rr}{\bm{r} }

\newcommand{\vv}{\bm{v} }
\newcommand{\xx}{\bm{x} }

\newcommand{\zz}{\bm{z} }

\newcommand{\eq}[1]{Eq.~\eqref{#1}}

\begin{document}

\preprint{XXX}

\title{Multipoint Turbulence Analysis with Helioswarm}

\author{Francesco Pecora}
\affiliation{Department of Physics and Astronomy, University of Delaware, Newark, DE 19716, USA}
\email{fpecora@udel.edu}

\author{Sergio Servidio}
\affiliation{Universit\`a della Calabria, Arcavacata di Rende, 87036, IT}

\author{Leonardo Primavera}
\affiliation{Universit\`a della Calabria, Arcavacata di Rende, 87036, IT}

\author{Antonella Greco}
\affiliation{Universit\`a della Calabria, Arcavacata di Rende, 87036, IT}

\author{Yan Yang}
\affiliation{Department of Physics and Astronomy, University of Delaware, Newark, DE 19716, USA}

\author{William H. Matthaeus}
\affiliation{Department of Physics and Astronomy, University of Delaware, Newark, DE 19716, USA}

\date{\today}

\begin{abstract}

Exploration of plasma dynamics in space, including turbulence, is entering a new era of multi-satellite constellation measurements that will determine fundamental properties with unprecedented precision.
Familiar but imprecise approximations will need to be abandoned and replaced with more advanced approaches. We present a preparatory study of the evaluation of second- and third-order statistics, using simultaneous measurements at many points.
Here, for specificity, the orbital configuration of the NASA Helioswarm mission is employed in conjunction with three-dimensional magnetohydrodynamics numerical simulations of turbulence. The Helioswarm 9-spacecraft constellation flies virtually through the turbulence to compare results with the exact numerical statistics. We demonstrate novel increment-based techniques for the computation of (1) the multidimensional spectra and (2) the turbulent energy flux. This latter increment-space estimate of the cascade rate, based on the third-order Yaglom-Politano-Pouquet theory, uses numerous increment-space tetrahedra. Our investigation reveals that Helioswarm will provide crucial information on the nature of astrophysical turbulence.

\end{abstract}

\maketitle



\textit{Introduction.--}Guidance from experiments remains a principal driver of progress in revealing the basic physics of turbulence, in spite of the difficulties inherent in diagnosing complex multiscale turbulent motions. Advances in this unsolved grand challenge problem immediately have beneficial impacts on numerous applications in space and astrophysical plasmas as well as geophysical fluids \cite{Pope,biskamp2003magnetohydrodynamic}. Laboratory turbulence experiments \cite{ComteBellotCorrsin71, YamadaEA06} have made great progress by employing numerous probes at multiple spatial positions.

In contrast, investigations of space plasma turbulence are typically limited to single spacecraft measurements, with a few notable exceptions. However, current state-of-the-art multispacecraft probes have severe limitations in quantifying interplanetary turbulence. The interspacecraft separations on the  Cluster \cite{credland1997cluster} mission are at a single scale, and too large for accurate computation of derivatives. The Magnetosphere Multiscale Mission (MMS) \cite{burch2016magnetospheric} probes very small sub-fluid scales, and  cannot accurately respond to  conditions in the ``pristine'' solar wind. The solution to these problems is, of course, a larger number of spacecraft, providing true multipoint multiscale measurements. The upcoming Helioswarm mission \cite{SpenceEA19} heralds several unprecedented advancements: nine spacecraft flying in the pristine solar wind, arranged such that the 36 baselines -- the separations between any two spacecraft -- range from a few tens to a thousand kilometers. This configuration allows computing derivatives with unrivaled precision and at several different scales centered on the turbulence inertial range. Here, we address the fundamental question of how to utilize such data from a turbulence theory perspective. Our conclusions impact not only Helioswarm but all future multipoint spacecraft constellations. The present work employs nominal orbital Helioswarm trajectories transferred in numerically generated turbulent fields, mimicking satellite flights through solar wind turbulence. The purpose is to propose novel methods to unambiguously characterize the inertial range of plasma turbulence. We base our new technique on multipoint increment analysis, extracting information about the spectra and the energy cascade rate of turbulence.

\textit{Numerical setup.--}We model decaying plasma turbulence by using magnetohydrodynamics (MHD) simulations, with and without mean magnetic field $B_0$ (along the $z$ axis). The simulations are carried out through a pseudo-spectral, incompressible, 3D numerical code that integrates the MHD equations in a three-periodic simulation box of 1024$^3$ gridpoints, having lengths in each direction equal to $2\pi L_0$, where we use classic Alfv\'enic units. The code uses a standard 2/3 dealiasing technique \cite{Orszag71f,Orszag72,OrszagTang79}. Both viscosity and resistivity are chosen to be adequately small, namely $\nu=\eta=5 \cdot 10^{-5}$. The initial conditions consist of a superposition of fluctuations with random phases in the range of modes peaked at $k=3$, with amplitude such that $v_\text{rms} = B_\text{rms} = 1$. This initial condition is evolved in time up to the peak of energy dissipation rate, which happens after a few Alfv\'en times. At this instant of time, turbulence is in a quasi-steady state \cite{Pouquet78, ServidioEA08-prld} and we perform our analysis.

\begin{figure}[ht]
    \centering
    \includegraphics[width=0.99\columnwidth]{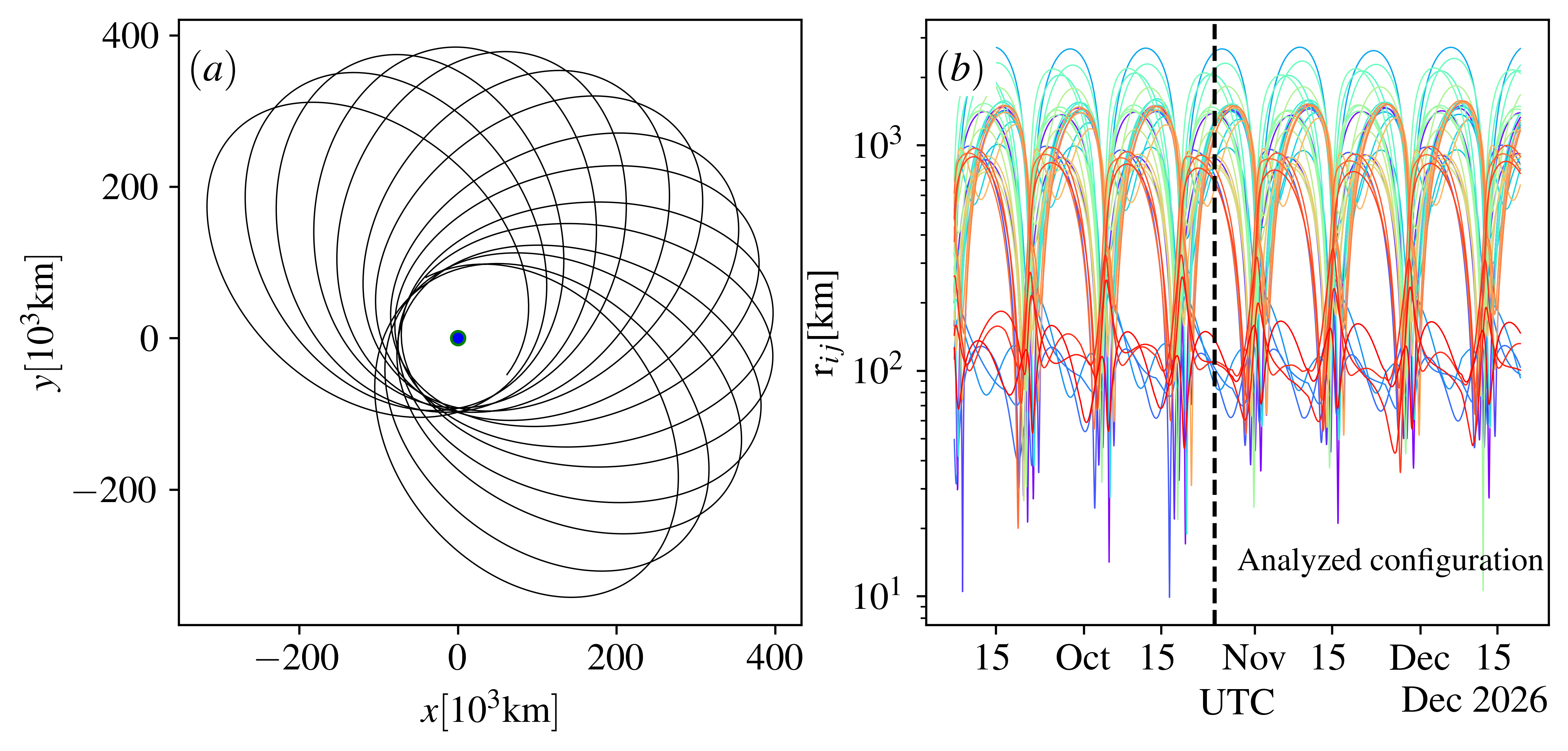}
    \caption{(a) Nominal phase Helioswarm trajectories projected in the ecliptic plane, with Earth indicated as a blue dot. (b) Interspacecraft separations. The vertical dashed line indicates one time at which Helioswarm separations are measured and transferred in the simulations (see text.)}
    \label{fig:ephemeris}
\end{figure}

\begin{figure}[ht]
    \centering
    \includegraphics[width=0.99\columnwidth]{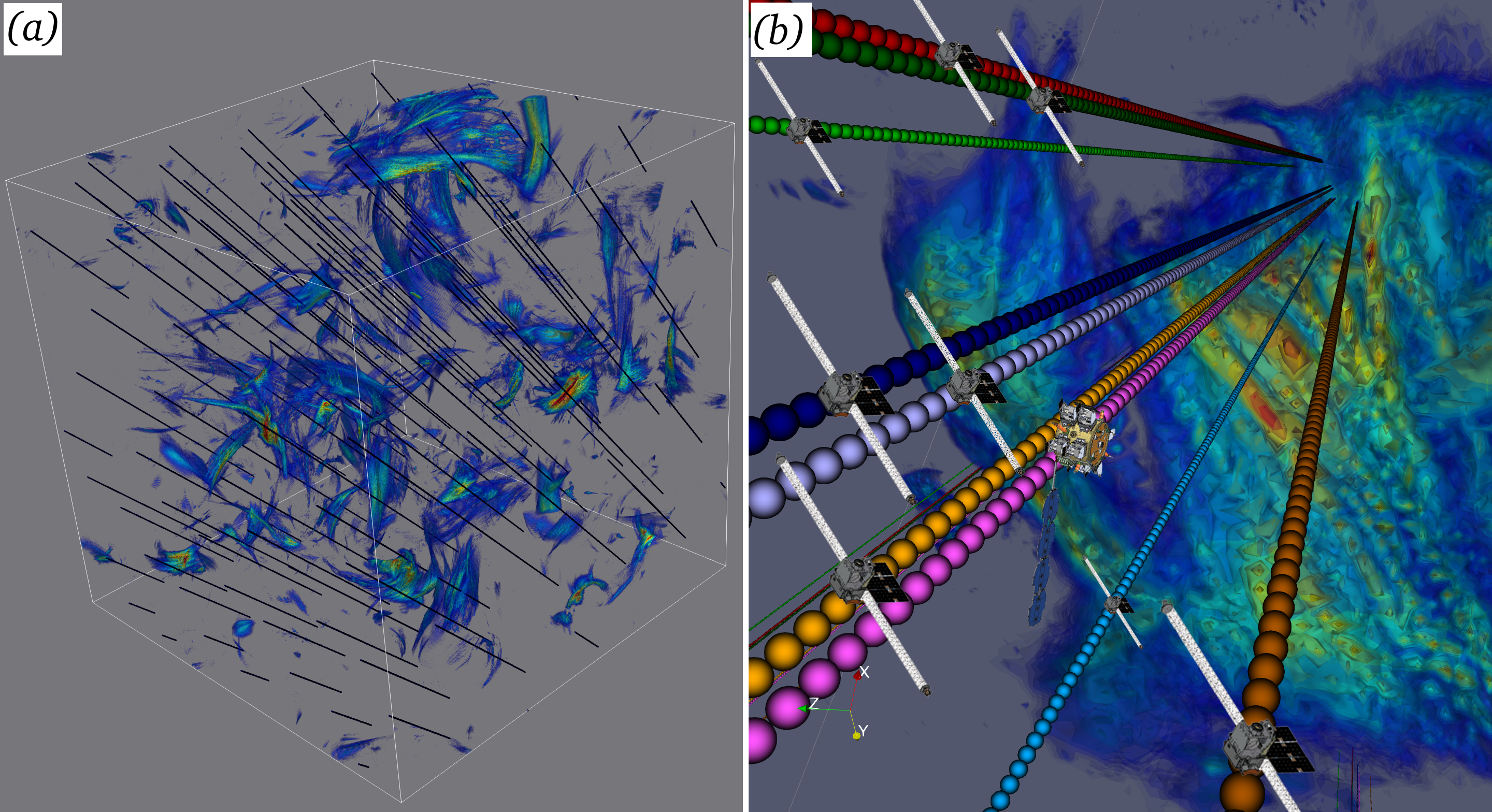}
    \caption{(a) Three-dimensional view of the simulation domain. Shaded colors are regions where the magnetic field is more intense. The black oblique lines are the virtual spacecraft trajectories. (b) Zoom into a region of very strong magnetic field, where individual spacecraft trajectories can be distinguished.}
    \label{fig:vsc_traj}
\end{figure}

\textit{Helioswarm trajectories.--}The nine-spacecraft constellation orbits the Earth, with a nominal period of two weeks, and with interspacecraft separations roughly ranging from 10 to 1000 km. The nominal phase trajectories projected onto the x-y plane, are shown in Fig.~\ref{fig:ephemeris}(a). The time evolution of the interspacecraft separations $r_{ij} = | \rr_i - \rr_j|$, where $i, j = 1, \dots, 9$, is shown in Fig.~\ref{fig:ephemeris}(b).
Knowing that the time evolution of the turbulence in the solar wind is much faster than the timescale at which the spacecraft  drift with respect to one another (few hours vs days), we can select and fix the separations at a single time, and then create virtual trajectories within the simulation volume of our MHD turbulence simulations. To do this, it is necessary to convert the relative positions of the spacecraft in numerical units to fit in the simulation domain. The conversion is made such that the minimal interspacecraft separation is set to be $50~\mbox{km}$ and then normalized to 10 times the Kolmogorov scale in the simulation. This normalization grants that the interspacecraft separations lie in the inertial range, as the Kolmogorov scale roughly indicates the smallest scale of the inertial range, and is defined as $\lambda_K = ( \nu^3 / \epsilon )^{1/4}$ \cite{politano1998dynamical} where $\nu$ is the kinematic viscosity and $\epsilon = \langle \eta j^2 + \nu \omega^2 \rangle$ is the total dissipation rate evaluated using the resistivity $\eta$, the current density $j$, and the vorticity $\omega$. With these assumptions, the trajectories are parallel lines that are then chosen to have a specified angle relative to the $z$ axis. As the virtual spacecraft motion progresses, the trajectories span the simulation box several times as shown by the black lines in Fig.~\ref{fig:vsc_traj}(a). The individual trajectories become visible when examined in zoomed-in regions, as shown by the dotted lines in Fig.~\ref{fig:vsc_traj}(b), where the shaded colors indicate a region of intense magnetic field. For the analyses that follow, the simulation data is interpolated onto the satellite trajectories.

The Helioswarm 9-spacecraft configuration allows different strategies for turbulence analyses based on increments, such as $\delta {\bf B}(\xx, \ellv) \equiv {\bf B}(\xx + \ellv) - {\bf B}(\xx)$, for the magnetic field ${\bf B}$, the position $\xx$ and the spatial lag $\ellv$. Particular examples are: (I) evaluation of increments at the fixed separations given by the 36 baselines. That is, $\BB(\xx)=\BB(\rr_i)$ and $\BB(\xx + \ellv )=\BB(\rr_j)$, where $i, j$ is any pair of spacecraft and the lag vector is $\ellv=\rr_{ij}$; (II) employing Taylor hypothesis, computing quantities along the 9 individual time series, i.e., $\BB(\xx)=\BB(\rr_i)$, $\BB(\xx + \ellv )=\BB(\rr_i-\boldsymbol{V}_{sw}\delta t)$ and $\ellv=-\boldsymbol{V}_{sw}\delta t$; (III) a combined scheme employing one spacecraft as a fixed point, and using the Taylor hypothesis, varying the lag relative to the paired partner \cite{OsmanHorbury07}:  $\BB(\xx)=\BB(\rr_i)$, $\BB(\xx + \ellv )=\BB(\rr_j-\boldsymbol{V}_{sw}\delta t)$ and $\ellv=\rr_{ij}-\boldsymbol{V}_{sw}\delta t$.

\textit{Power spectra.--}We first carry out an increment-space estimation of the power spectra. The method relies on the Blackman-Tukey technique estimation of the second-order structure function \cite{matthaeus1982measurement}, after which the magnetic field power spectral density (PSD) is obtained via Fourier transform of the autocorrelation function. Except for the total variance, the autocorrelation is readily obtained from the second-order structure function, as described below. We proceed with the analysis of the magnetic field, but the same procedure can be applied to the density and fluid velocity. The magnetic field second-order structure function is defined as $S^2_b( \ellv ) = \langle | \BB(\xx) - \BB(\xx + \ellv ) |^2 \rangle$, where the averaging operation $\langle \cdot \rangle$ is performed over a suitable volume. 

We have employed the above procedure (structure function, autocorrelation function, and Blackman-Tukey spectrum) to obtain second-order turbulence statistics using strategy III. We supplement this technique with repetitive passage through the simulation with varying angular orientations of the trajectories relative to the box. The latter procedure emulates analyzing solar wind streams with the mean field being oriented in different directions. When the mean field is absent, the procedure gives a more ergodic sampling of turbulence. Generally, we find good correspondence between the methods. For example, the second-order structure functions obtained with strategy I (not shown) produce 36 points nicely scattered about the globally computed structure function for the isotropic case.

\begin{figure}[ht]
    \centering
    \includegraphics[width=0.99\columnwidth]{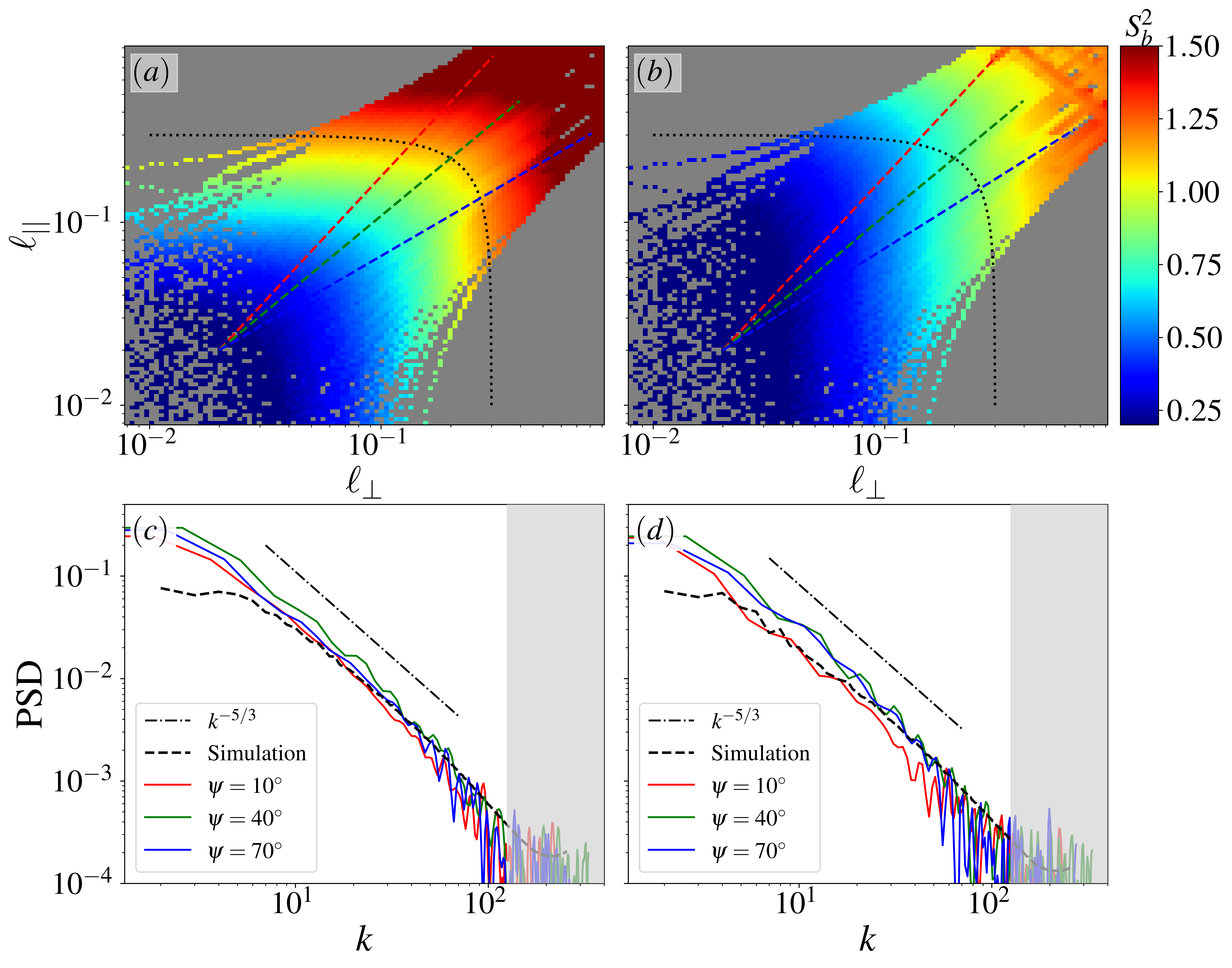}
    \caption{(Top) Structure functions (strategy III) for (a) isotropic and (b) anisotropic simulations in the parallel-perpendicular increment plane. Dashed colored lines are directions along which 1D cuts are collected. The dotted line indicates the correlation length (of the isotropic case). (Bottom) PSDs obtained from 1D cuts compared with exact PSD obtained from simulation (dashed line), for isotropic (c) and anisotropic (d) cases.}
    \label{fig:s2b_9SC}
\end{figure}

To populate correlations in a plane of parallel and perpendicular increments, we generated several different sets of Helioswarm-like trajectories changing the angle with respect to the $z$ axis: When the angle is smaller, more parallel coverage is obtained. Accordingly, for angles closer to $\pi/2$, more perpendicular coverage is realized. We merged 6 different trajectory inclinations from $20^\circ$ to $70^\circ$. The structure function in the perpendicular-parallel plane is shown in Fig.~\ref{fig:s2b_9SC}(a,~b). 
The 2D structure functions clearly show the effect of a mean field, the contours of the structure function being squashed in the perpendicular direction for the anisotropic case. We can further look at directional properties by sampling the 2D structure functions along 1D cuts in different directions. As expected, the 1D cuts in the isotropic case show no differences. On the other hand, in the anisotropic case, appreciable differences arise along different directions relative to the mean field. 

The directional dependence of the structure function translates directly into the directional anisotropy of the magnetic field power spectrum. Indeed, the structure function and the correlation functions are related as $C(\ellv) = E_b - \frac{1}{2} S^2_b( \ellv )$, where $E_b = \langle \delta b^2 \rangle$ is (twice) the energy density of the magnetic fluctuations. This link between $S^2_b$ and $C$ is important for at least two reasons: (I) the Fourier transform relates the correlation function to the power spectrum, and (II) the structure function has stronger convergence properties than the correlation function. 

Starting from $S^2_b$, it is possible to recover $C$, then Fourier transform $C$ and obtain the power spectrum. However, some care needs to be exercised. The correlation function must be an even function of the lag. This and formal periodicity properties are prescribed by reflecting the correlation function about the origin. To avoid spurious oscillations at large lags due to low statistical weight, the correlation function is windowed with a cosine function that smooths the far edges gently to zero. We zero-pad the correlation function to extend the domain without adding any further information, with the aesthetic advantage of better-resolving modes to the inertial range \cite{matthaeus1982measurement}. Finally, the Fourier transform of the assembled correlation function yields the power spectrum of the magnetic field \cite{batchelor1953theory}.

In Fig.~\ref{fig:s2b_9SC}, the power spectra computed from the 1D cuts are shown for the (c) isotropic, and (d) anisotropic cases. These are shown together with the exact isotropic spectrum obtained directly from the full simulation dataset. 
Several features can now be noticed: 
(I) different extension in $k$ between simulation and directional spectra. For the latter, larger k's appear because of a finer sampling of the second-order structure functions when collecting 1D information. Smaller k's arise based on the total length of the trajectories ``time series'' that depends on how many times the trajectories sample the simulation domain.
(II) In fact, the gray shaded area in panels (c) and (d) identifies k values related to separations smaller than the smallest spacecraft separation (that is $10 \lambda_K$).
(III) The slope in the inertial range is overall consistent between all the different spectra (being isotropic, anisotropic, and exact). 
(IV) In the isotropic case, the spectral modes' magnitudes remain nearly constant regardless of the sampling direction.
(V) In the anisotropic case, the nearly parallel $10^\circ$ spectrum is of smaller magnitudes than the spectra at more oblique directions, and also has, one may argue, a shorter inertial range.
(VI) In passing, it is interesting to notice that, despite the structure functions not showing a neat inertial range (not shown here), their Fourier transforms (the spectra) do. This kind of analysis is of fundamental importance in order to predict what can be observed with constellations such as Helioswarm in the solar wind, where the exact spectrum is not available for comparison.

\textit{Energy cascade rate.--}The energy cascade rate is a fundamental ingredient of turbulence theory, and below we measure it with a novel technique. Numerous attempts have been made to estimate this number in space plasmas 
 \cite{sorrisovalvo2018statistical, bandyopadhyay2020situ, MarinoSorriso23}. However, the lack of multipoint measurements in the appropriate environment or range of scales has made it necessary to rely on various simplifying approximations that may provide potentially unrealistic estimates (see, e.g., \cite{WangEA22}). In the incompressible regime, the cascade rate $\epsilon$ is related to the increments of the Els\"asser variables via the von K\'arm\'an-Howarth equations $\frac{\partial}{\partial t}\langle |\delta \mathbf{z}^\pm|^2 \rangle = - \mathbf{\nabla}_{\boldsymbol{\ell}} \cdot \langle \delta \mathbf{z}^\mp |\delta \mathbf{z}^\pm|^2 \rangle + 2 \nu \nabla^2_{\boldsymbol{\ell}}\langle |\delta \mathbf{z}^\pm|^2 \rangle - 4\epsilon^\pm$ where, $\delta \mathbf{z}^\pm (\mathbf{x}, \boldsymbol{\ell}) = \mathbf{z}^\pm (\mathbf{x} + \boldsymbol{\ell}) - \mathbf{z}^\pm (\mathbf{x})$ are the increments of the Els\"asser variables $\zz^\pm~=~\vv~\pm~\bb$. Here $\vv$ and $\bb$ are the velocity and magnetic fields respectively, and the magnetic field is in Alfv\'en speed units. The averaging operation $\langle \cdot \rangle$ is performed over a suitably large domain in real space. These equations are exact for homogeneous turbulence, at any lag $\boldsymbol{\ell}$. 

For a large, scale-separated system, the different terms separately dominate at different length-scales: Generally, the time derivative is large at very large scales, the dissipative term is large at very small scales, while the non-linear term, also called the Yaglom term, dominates in the intermediate inertial range. Therefore, when one focuses on the inertial range, the full von K\'arm\'an-Howarth equation reduces to the Yaglom law \cite{PolitanoEA98}, 
\begin{equation}
    \mathbf{\nabla}_{\boldsymbol{\ell}} \cdot \mathbf{Y^\pm} = -4 \epsilon^\pm, 
\label{eq:Yaglom}
\end{equation}
that involves only the third-order structure function (or Yaglom flux) $\mathbf{Y^\pm} = \langle \delta \mathbf{z}^\mp |\delta \mathbf{z}^\pm|^2 \rangle$. The dissipation rate is finally given by $\epsilon=(\epsilon^+ +\epsilon^-)/2$.

In the pre-Helioswarm era, even this simpler reduced form of the cascade law was relatively inaccessible via spacecraft measurements for the lack of viable multipoint measurements. Attempts were made with Cluster \cite{OsmanEA11-3rd} and MMS \cite{BandyopadhyayEA20-PSP3rd} but these are obviously limited to four points and intrinsically to a single interspacecraft scale. Helioswarm \cite{SpenceEA19} introduces a novel configuration of 9 spacecraft that provides 36 baselines, most of which will lie in the inertial range where Eq.~\ref{eq:Yaglom} is valid. It is immediately evident that the available data to solve Eq.~\ref{eq:Yaglom} is a steeply increasing function of the number of simultaneous measurement points. The 36 baselines are geometric lines in the real space and become 36 points in the lag space where the divergence is to be computed. This means that, in lag space, we have a swarm of 36 points, at each of which we have a value of $\YY^\pm$. 
\begin{figure}[ht!]
    \centering
    \includegraphics[width=0.99\columnwidth]{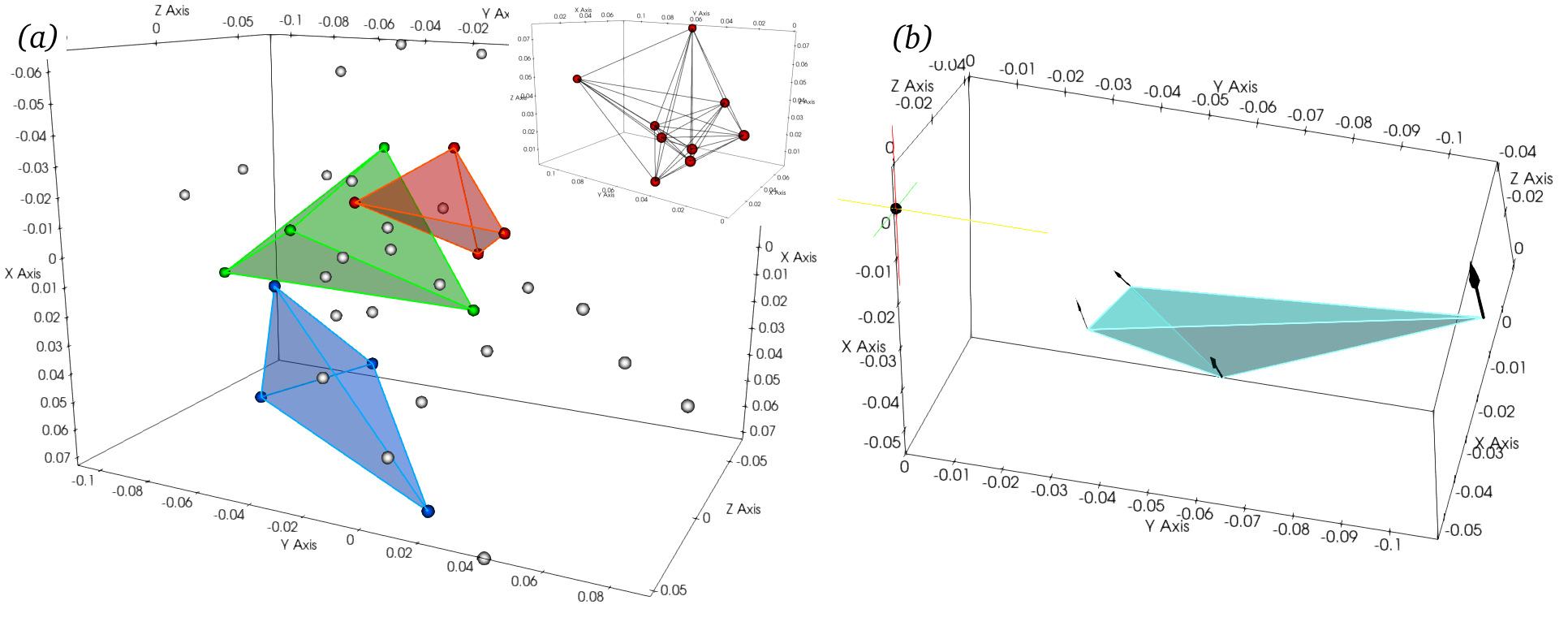}
    \caption{(inset) Nine spacecraft (red spheres) provide 36 baselines (black lines) that correspond to 36 points in lag space (panel a, spheres). Three possible tetrahedra are highlighted in colors. (b) Tetrahedron in lag space with the Yaglom flux vectors at its vertices pointing roughly towards the origin (black dot on the left). Arrow length is $\propto |\YY|$.}
    \label{fig:vesparo}
\end{figure}
Implementing the new approach, we sort the 36 points in permutations of 4 to form the astonishing number of 58905 tetrahedra (of which, we used 56718). To compute the required lag-space divergence, the tetrahedra are subjected to well-tested techniques based on the curlometer approach \cite{dunlop2002four}, that have been developed to analyze Cluster and MMS data. This procedure is explicated in Fig.~\ref{fig:vesparo} where the 9 spacecraft  are represented in real space (inset) as (red) spheres with the 36 baselines drawn in black. Panel (a) of Fig.~\ref{fig:vesparo} shows the lag space, where the baselines transform into 36 points -- represented as spheres --, and 3 sample tetrahedra are color shaded. Note that only strategy I is used here to evaluate increments and only a single realization of the trajectories is employed.

Panel (b), instead, depicts a tetrahedron in lag space, where at each vertex the Yaglom vector $\mathbf{Y}= ( \mathbf{Y^+} + \mathbf{Y^-} ) / 2$ is represented with length proportional to its magnitude. The arrows do, indeed, point roughly toward the origin (black dot) with decreasing magnitude moving towards smaller scales in the expected fashion \cite{verdini2015anisotropy}. This is expected from the general structure of Eq.~\ref{eq:Yaglom}, from which, for $\epsilon^\pm \sim$ constant, the Yaglom flux is expected to be $\sim \ell$.
\begin{figure}[ht]
    \centering
    \includegraphics[width=0.99\columnwidth]{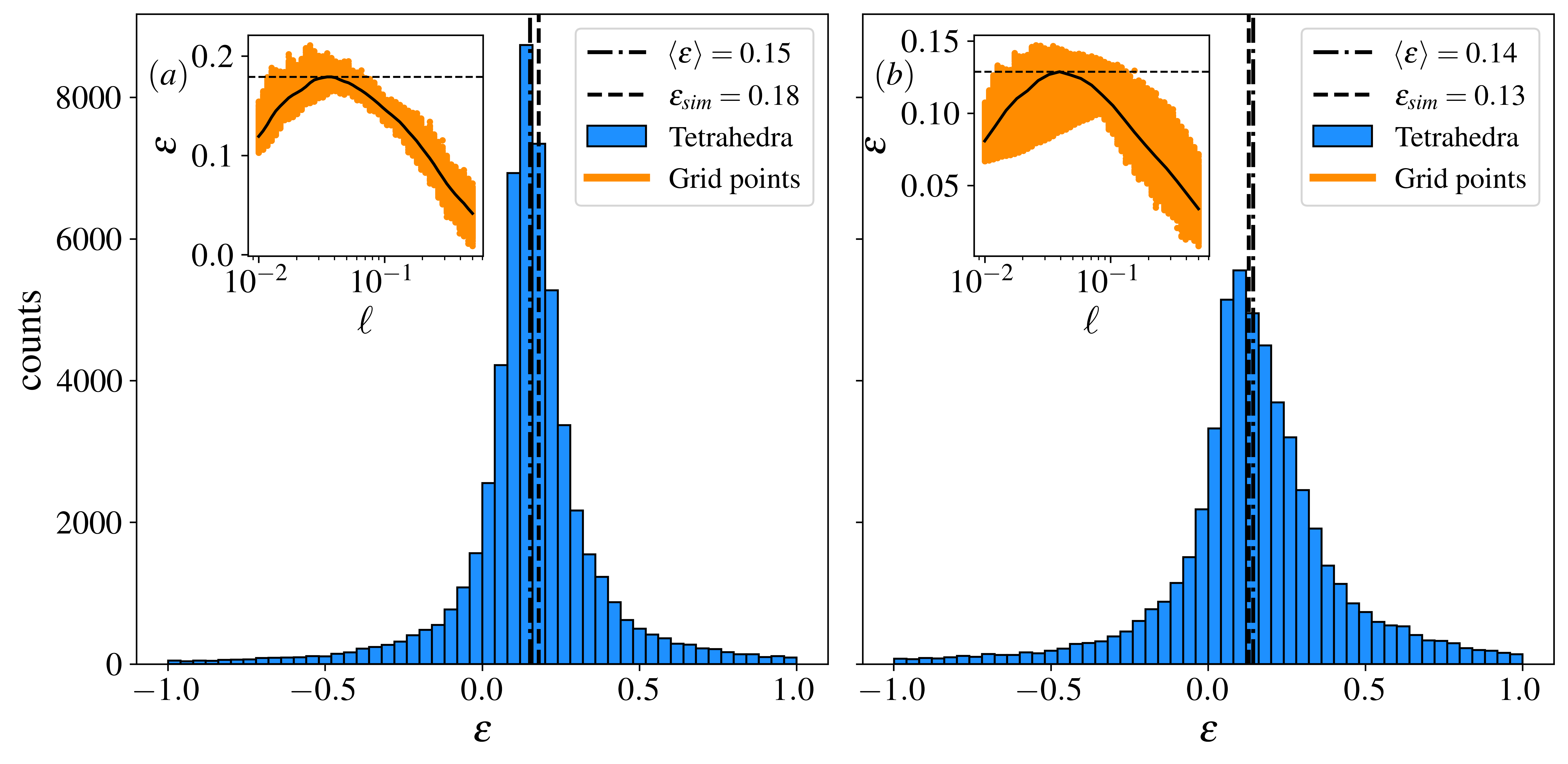}
    \caption{Estimations of cascade rate, for isotropic (a) and anisotropic (b) simulations. Insets show $\epsilon(\ell, \theta, \phi)$ computed from Yaglom's law \eq{eq:Yaglom} over all  grid points, plotted as a function of lag. The maximum value within the inertial range is highlighted with a dashed horizontal line and reported as $\epsilon_{sim}$ in the legend. Histograms are obtained from computing the divergence using tetrahedra in lag space (see text and Fig.~\ref{fig:vesparo}). We discarded values larger than $\pm 1$.}
    \label{fig:divY}
\end{figure}
We start by computing Yaglom's law, Eq.~\ref{eq:Yaglom} in the simulation using spherical coordinates and all simulation grid points, to have the exact value as a reference. The insets in Fig.~\ref{fig:divY} show the values of $\epsilon(\ell,\theta,\phi)$ as a function of the lag; Panel~(a) and (b) are for the isotropic and anisotropic cases, respectively. The variability of $\epsilon(\ell,\theta,\phi)$ at each lag $\ell$, for varying $\theta$ and $\phi$ (the azimuthal and polar angles in the simulation domain), is indicated by the spread around the mean value (black curve). These variations are attributed to inhomogeneities and anisotropies (the latter is evidently more present in the simulation with the mean field). The maximum of the average curve is identified as the ``effective'' cascade rate:  $\epsilon_{sim} = 0.18$ and $\epsilon_{sim} = 0.13$ for the isotropic and anisotropic simulations, respectively, which also indicates the points where the inertial range conditions are better attained. 

The histograms in Fig.~\ref{fig:divY} represent the values of $\epsilon$ measured by calculating the divergence over the tetrahedra in lag space. The effective and averaged values of $\epsilon$ are shown as vertical lines. The agreement between the cascade rate obtained using the tetrahedra with the exact one is excellent as the relative errors are 7\%  and 15\% for the anisotropic and isotropic cases, respectively.

\textit{Discussion.--} We have shown that sampling data from many spacecraft in a realistic constellation orbit can accurately describe statistics based on increments, including second and third orders statistics. These crucially lead to the detection of anisotropy, accurate inertial range spectra estimate, and perhaps most importantly, the evaluation of the turbulence energy transfer rate in the inertial range. This is accomplished here for the first time using nine-point sampling and nominal Helioswarm orbits, scale-adjusted to a high-resolution MHD turbulence simulation. The accurate evaluation of the cascade rate is based on a novel strategy in which the von K\'arm\'an-Yaglom expression for the cascade rate is solved over the more than 56,000 lag-space tetrahedra provided by the Helioswarm constellation. The statistical agreement between the cascade rate measured using the tetrahedra with that obtained from the exact evaluation over the grid points is striking and very promising for the ability of Helioswarm to determine accurate approximations of the solar wind turbulence cascade rate. We expect that improved estimates will be obtained by applying geometric quality factors \cite{dunlop2002four} to the tetrahedra. Another detail that can be observed, is that the histogram of $\epsilon$ estimates in the anisotropic case reflects the variability of values in different directions where the inertial ranges may have different extensions as shown by the spectra in Fig.~\ref{fig:s2b_9SC}. Further experience with analysis of cascade rates in simulations \cite{WangEA22} will guide refinements of this method, as well as extensions to properly define and obtain directional cascade rates as well.

Note that the span of scale in the solar wind is much larger than that available in the simulation. However, only the largest virtual spacecraft separations approach or exceed the scales within the inertial range. Since our goal here is to evaluate inertial range statistics, the results are not  severely affected. When applied to the magnetosheath \cite{bandyopadhyay2018incompressive} the range of scales is narrower and an even closer correspondence can be found with the simulation. 

We note that higher-order polyhedra with vertices $>4$ might also be employed. For this work, we used tetrahedra to exploit well-tested routines that have been validated in Cluster and MMS data \cite{dunlop2002four}. The present demonstration provides guidance and confidence concerning the evaluation of critical turbulence quantities on the upcoming generation of multispacecraft constellations beyond Helioswarm, including concepts such as MagCon, Plasma Observatory, and Magnetore \cite{maruca2021magnetore}. Accurate evaluations of cascade rates directly support theories of dissipation, plasma heating, solar wind acceleration, and cross-scale dynamics in general, which through these missions may well revolutionize our conception of the dynamics in these complex interplanetary and magnetosphere space plasmas. \\

\begin{acknowledgments}
This research is supported in part by the MMS Theory and Modeling program grant 80NSSC19K0284,  the Parker Solar prove Guest Investigator program 80NSSC21K1765, the PUNCH mission through SWRI subcontract N99054DS, and the NSF/DOE program under grant AGS-2108834 at the University of Delaware.
LP acknowledges support by EU FP7 2007-13 through the MATERIA Project (PONa3\_00370) and EU Horizon 2020 through the STAR\_2 Project (PON R\&I 2014-20, PIR01\_00008) for running the simulations on the ``newton'' cluster.
Helioswarm sprites in Fig.~\ref{fig:vsc_traj} courtesy of UNH \protect\url{https://eos.unh.edu/helioswarm/mission/flight-system}.
\end{acknowledgments}

\bibliography{biblio_ciccio,ag,hl,mp,qz,refs_WHM}

 \newcommand{\BIBand} {and} 
  \newcommand{\boldVol}[1] {\textbf{#1}} 
  \providecommand{\SortNoop}[1]{} 
  \providecommand{\sortnoop}[1]{} 
  \newcommand{\stereo} {\emph{{S}{T}{E}{R}{E}{O}}} 
  \newcommand{\au} {{A}{U}\ } 
  \newcommand{\AU} {{A}{U}\ } 
  \newcommand{\MHD} {{M}{H}{D}\ } 
  \newcommand{\mhd} {{M}{H}{D}\ } 
  \newcommand{\RMHD} {{R}{M}{H}{D}\ } 
  \newcommand{\rmhd} {{R}{M}{H}{D}\ } 
  \newcommand{\wkb} {{W}{K}{B}\ } 
  \newcommand{\alfven} {{A}lfv{\'e}n\ } 
  \newcommand{\alfvenic} {{A}lfv{\'e}nic\ } 
  \newcommand{\Alfven} {{A}lfv{\'e}n\ } 
  \newcommand{\Alfvenic} {{A}lfv{\'e}nic\ }
\begin{thebibliography}{27}%
\makeatletter
\providecommand \@ifxundefined [1]{%
 \@ifx{#1\undefined}
}%
\providecommand \@ifnum [1]{%
 \ifnum #1\expandafter \@firstoftwo
 \else \expandafter \@secondoftwo
 \fi
}%
\providecommand \@ifx [1]{%
 \ifx #1\expandafter \@firstoftwo
 \else \expandafter \@secondoftwo
 \fi
}%
\providecommand \natexlab [1]{#1}%
\providecommand \enquote  [1]{``#1''}%
\providecommand \bibnamefont  [1]{#1}%
\providecommand \bibfnamefont [1]{#1}%
\providecommand \citenamefont [1]{#1}%
\providecommand \href@noop [0]{\@secondoftwo}%
\providecommand \href [0]{\begingroup \@sanitize@url \@href}%
\providecommand \@href[1]{\@@startlink{#1}\@@href}%
\providecommand \@@href[1]{\endgroup#1\@@endlink}%
\providecommand \@sanitize@url [0]{\catcode `\\12\catcode `\$12\catcode
  `\&12\catcode `\#12\catcode `\^12\catcode `\_12\catcode `\%12\relax}%
\providecommand \@@startlink[1]{}%
\providecommand \@@endlink[0]{}%
\providecommand \url  [0]{\begingroup\@sanitize@url \@url }%
\providecommand \@url [1]{\endgroup\@href {#1}{\urlprefix }}%
\providecommand \urlprefix  [0]{URL }%
\providecommand \Eprint [0]{\href }%
\providecommand \doibase [0]{https://doi.org/}%
\providecommand \selectlanguage [0]{\@gobble}%
\providecommand \bibinfo  [0]{\@secondoftwo}%
\providecommand \bibfield  [0]{\@secondoftwo}%
\providecommand \translation [1]{[#1]}%
\providecommand \BibitemOpen [0]{}%
\providecommand \bibitemStop [0]{}%
\providecommand \bibitemNoStop [0]{.\EOS\space}%
\providecommand \EOS [0]{\spacefactor3000\relax}%
\providecommand \BibitemShut  [1]{\csname bibitem#1\endcsname}%
\let\auto@bib@innerbib\@empty
\bibitem [{\citenamefont {Pope}(2000)}]{Pope}%
  \BibitemOpen
  \bibfield  {author} {\bibinfo {author} {\bibfnamefont {S.~B.}\ \bibnamefont
  {Pope}},\ }\href@noop {} {\emph {\bibinfo {title} {Turbulent Flows}}}\
  (\bibinfo  {publisher} {Cambridge University Press},\ \bibinfo {address}
  {Cambridge, UK},\ \bibinfo {year} {2000})\BibitemShut {NoStop}%
\bibitem [{\citenamefont {Biskamp}(2003)}]{biskamp2003magnetohydrodynamic}%
  \BibitemOpen
  \bibfield  {author} {\bibinfo {author} {\bibfnamefont {D.}~\bibnamefont
  {Biskamp}},\ }\href@noop {} {\emph {\bibinfo {title} {Magnetohydrodynamic
  turbulence}}}\ (\bibinfo  {publisher} {Cambridge University Press},\ \bibinfo
  {year} {2003})\BibitemShut {NoStop}%
\bibitem [{\citenamefont {{Comte-Bellot}}\ and\ \citenamefont
  {{Corrsin}}(1971)}]{ComteBellotCorrsin71}%
  \BibitemOpen
  \bibfield  {author} {\bibinfo {author} {\bibfnamefont {G.}~\bibnamefont
  {{Comte-Bellot}}}\ and\ \bibinfo {author} {\bibfnamefont {S.}~\bibnamefont
  {{Corrsin}}},\ }\bibfield  {title} {\bibinfo {title} {{Simple Eulerian time
  correlation of full- and narrow-band velocity signals in grid-generated,
  `isotropic' turbulence}},\ }\href {https://doi.org/10.1017/S0022112071001599}
  {\bibfield  {journal} {\bibinfo  {journal} {Journal of Fluid Mechanics}\
  }\textbf {\bibinfo {volume} {48}},\ \bibinfo {pages} {273} (\bibinfo {year}
  {1971})}\BibitemShut {NoStop}%
\bibitem [{\citenamefont {{Yamada}}\ \emph {et~al.}(2006)\citenamefont
  {{Yamada}}, \citenamefont {{Ren}}, \citenamefont {{Ji}}, \citenamefont
  {{Breslau}}, \citenamefont {{Gerhardt}}, \citenamefont {{Kulsrud}},\ and\
  \citenamefont {{Kuritsyn}}}]{YamadaEA06}%
  \BibitemOpen
  \bibfield  {author} {\bibinfo {author} {\bibfnamefont {M.}~\bibnamefont
  {{Yamada}}}, \bibinfo {author} {\bibfnamefont {Y.}~\bibnamefont {{Ren}}},
  \bibinfo {author} {\bibfnamefont {H.}~\bibnamefont {{Ji}}}, \bibinfo {author}
  {\bibfnamefont {J.}~\bibnamefont {{Breslau}}}, \bibinfo {author}
  {\bibfnamefont {S.}~\bibnamefont {{Gerhardt}}}, \bibinfo {author}
  {\bibfnamefont {R.}~\bibnamefont {{Kulsrud}}},\ and\ \bibinfo {author}
  {\bibfnamefont {A.}~\bibnamefont {{Kuritsyn}}},\ }\bibfield  {title}
  {\bibinfo {title} {{Experimental study of two-fluid effects on magnetic
  reconnection in a laboratory plasma with variable collisionality}},\ }\href
  {https://doi.org/10.1063/1.2203950} {\bibfield  {journal} {\bibinfo
  {journal} {Physics of Plasmas}\ }\textbf {\bibinfo {volume} {13}},\ \bibinfo
  {eid} {052119} (\bibinfo {year} {2006})}\BibitemShut {NoStop}%
\bibitem [{\citenamefont {Credland}\ \emph {et~al.}(1997)\citenamefont
  {Credland}, \citenamefont {Mecke},\ and\ \citenamefont
  {Ellwood}}]{credland1997cluster}%
  \BibitemOpen
  \bibfield  {author} {\bibinfo {author} {\bibfnamefont {J.}~\bibnamefont
  {Credland}}, \bibinfo {author} {\bibfnamefont {G.}~\bibnamefont {Mecke}},\
  and\ \bibinfo {author} {\bibfnamefont {J.}~\bibnamefont {Ellwood}},\
  }\bibfield  {title} {\bibinfo {title} {The cluster mission: Esa's spacefleet
  to the magnetosphere},\ }\href {https://doi.org/10.1023/A:1004914822769}
  {\bibfield  {journal} {\bibinfo  {journal} {Space Science Reviews}\ }\textbf
  {\bibinfo {volume} {79}},\ \bibinfo {pages} {33} (\bibinfo {year}
  {1997})}\BibitemShut {NoStop}%
\bibitem [{\citenamefont {Burch}\ \emph {et~al.}(2016)\citenamefont {Burch},
  \citenamefont {Moore}, \citenamefont {Torbert},\ and\ \citenamefont
  {Giles}}]{burch2016magnetospheric}%
  \BibitemOpen
  \bibfield  {author} {\bibinfo {author} {\bibfnamefont {J.~L.}\ \bibnamefont
  {Burch}}, \bibinfo {author} {\bibfnamefont {T.~E.}\ \bibnamefont {Moore}},
  \bibinfo {author} {\bibfnamefont {R.~B.}\ \bibnamefont {Torbert}},\ and\
  \bibinfo {author} {\bibfnamefont {B.~L.}\ \bibnamefont {Giles}},\ }\bibfield
  {title} {\bibinfo {title} {Magnetospheric multiscale overview and science
  objectives},\ }\href {https://doi.org/10.1007/s11214-015-0164-9} {\bibfield
  {journal} {\bibinfo  {journal} {Space Science Reviews}\ }\textbf {\bibinfo
  {volume} {199}},\ \bibinfo {pages} {5} (\bibinfo {year} {2016})}\BibitemShut
  {NoStop}%
\bibitem [{\citenamefont {{Spence}}(2019)}]{SpenceEA19}%
  \BibitemOpen
  \bibfield  {author} {\bibinfo {author} {\bibfnamefont {H.~E.}\ \bibnamefont
  {{Spence}}},\ }\bibfield  {title} {\bibinfo {title} {{HelioSwarm: Unlocking
  the Multiscale Mysteries of Weakly-Collisional Magnetized Plasma Turbulence
  and Ion Heating}},\ }in\ \href@noop {} {\emph {\bibinfo {booktitle} {AGU Fall
  Meeting Abstracts}}},\ Vol.\ \bibinfo {volume} {2019}\ (\bibinfo {year}
  {2019})\ pp.\ \bibinfo {pages} {SH11B--04}\BibitemShut {NoStop}%
\bibitem [{\citenamefont {Orszag}(1971)}]{Orszag71f}%
  \BibitemOpen
  \bibfield  {author} {\bibinfo {author} {\bibfnamefont {S.~A.}\ \bibnamefont
  {Orszag}},\ }\bibfield  {title} {\bibinfo {title} {On the elimination of
  aliasing in finite-difference schemes by filtering high-wavenumber
  components},\ }\href
  {https://doi.org/10.1175/1520-0469(1971)028<1074:OTEOAI>2.0.CO;2} {\bibfield
  {journal} {\bibinfo  {journal} {J.\ Atmos.\ Sci.}\ }\textbf {\bibinfo
  {volume} {28}},\ \bibinfo {pages} {1074} (\bibinfo {year}
  {1971})}\BibitemShut {NoStop}%
\bibitem [{\citenamefont {Orszag}(1972)}]{Orszag72}%
  \BibitemOpen
  \bibfield  {author} {\bibinfo {author} {\bibfnamefont {S.~A.}\ \bibnamefont
  {Orszag}},\ }\bibfield  {title} {\bibinfo {title} {Comparison of
  pseudospectral and spectral approximation},\ }\href@noop {} {\bibfield
  {journal} {\bibinfo  {journal} {Stud. Applied Math.}\ }\textbf {\bibinfo
  {volume} {51}},\ \bibinfo {pages} {253} (\bibinfo {year} {1972})}\BibitemShut
  {NoStop}%
\bibitem [{\citenamefont {Orszag}\ and\ \citenamefont
  {Tang}(1979)}]{OrszagTang79}%
  \BibitemOpen
  \bibfield  {author} {\bibinfo {author} {\bibfnamefont {S.~A.}\ \bibnamefont
  {Orszag}}\ and\ \bibinfo {author} {\bibfnamefont {C.-M.}\ \bibnamefont
  {Tang}},\ }\bibfield  {title} {\bibinfo {title} {Small-scale structure of
  two-dimensional magnetohydrodynamic turbulence},\ }\href@noop {} {\bibfield
  {journal} {\bibinfo  {journal} {J.\ Fluid Mech.}\ }\textbf {\bibinfo {volume}
  {90}},\ \bibinfo {pages} {129} (\bibinfo {year} {1979})}\BibitemShut
  {NoStop}%
\bibitem [{\citenamefont {Pouquet}(1978)}]{Pouquet78}%
  \BibitemOpen
  \bibfield  {author} {\bibinfo {author} {\bibfnamefont {A.}~\bibnamefont
  {Pouquet}},\ }\bibfield  {title} {\bibinfo {title} {On two-dimensional
  magnetohydrodynamic turbulence},\ }\href
  {https://doi.org/10.1017/S0022112078001950} {\bibfield  {journal} {\bibinfo
  {journal} {J. Fluid Mech.}\ }\textbf {\bibinfo {volume} {88}},\ \bibinfo
  {pages} {1} (\bibinfo {year} {1978})}\BibitemShut {NoStop}%
\bibitem [{\citenamefont {{Servidio}}\ \emph {et~al.}(2008)\citenamefont
  {{Servidio}}, \citenamefont {{Matthaeus}},\ and\ \citenamefont
  {{Dmitruk}}}]{ServidioEA08-prld}%
  \BibitemOpen
  \bibfield  {author} {\bibinfo {author} {\bibfnamefont {S.}~\bibnamefont
  {{Servidio}}}, \bibinfo {author} {\bibfnamefont {W.~H.}\ \bibnamefont
  {{Matthaeus}}},\ and\ \bibinfo {author} {\bibfnamefont {P.}~\bibnamefont
  {{Dmitruk}}},\ }\bibfield  {title} {\bibinfo {title} {{Depression of
  Nonlinearity in Decaying Isotropic MHD Turbulence}},\ }\href
  {https://doi.org/10.1103/PhysRevLett.100.095005} {\bibfield  {journal}
  {\bibinfo  {journal} {Physical Review Letters}\ }\textbf {\bibinfo {volume}
  {100}},\ \bibinfo {pages} {095005} (\bibinfo {year} {2008})}\BibitemShut
  {NoStop}%
\bibitem [{\citenamefont {Politano}\ and\ \citenamefont
  {Pouquet}(1998)}]{politano1998dynamical}%
  \BibitemOpen
  \bibfield  {author} {\bibinfo {author} {\bibfnamefont {H.}~\bibnamefont
  {Politano}}\ and\ \bibinfo {author} {\bibfnamefont {A.}~\bibnamefont
  {Pouquet}},\ }\bibfield  {title} {\bibinfo {title} {Dynamical length scales
  for turbulent magnetized flows},\ }\href
  {https://doi.org/https://doi.org/10.1029/97GL03642} {\bibfield  {journal}
  {\bibinfo  {journal} {Geophysical Research Letters}\ }\textbf {\bibinfo
  {volume} {25}},\ \bibinfo {pages} {273} (\bibinfo {year} {1998})}\BibitemShut
  {NoStop}%
\bibitem [{\citenamefont {Osman}\ and\ \citenamefont
  {Horbury}(2007)}]{OsmanHorbury07}%
  \BibitemOpen
  \bibfield  {author} {\bibinfo {author} {\bibfnamefont {K.~T.}\ \bibnamefont
  {Osman}}\ and\ \bibinfo {author} {\bibfnamefont {T.~S.}\ \bibnamefont
  {Horbury}},\ }\bibfield  {title} {\bibinfo {title} {Multispacecraft
  measurement of anisotropic correlation functions in solar wind turbulence},\
  }\href@noop {} {\bibfield  {journal} {\bibinfo  {journal} {Astrophys.\ J.}\
  }\textbf {\bibinfo {volume} {654}},\ \bibinfo {pages} {L103} (\bibinfo {year}
  {2007})}\BibitemShut {NoStop}%
\bibitem [{\citenamefont {Matthaeus}\ and\ \citenamefont
  {Goldstein}(1982)}]{matthaeus1982measurement}%
  \BibitemOpen
  \bibfield  {author} {\bibinfo {author} {\bibfnamefont {W.~H.}\ \bibnamefont
  {Matthaeus}}\ and\ \bibinfo {author} {\bibfnamefont {M.~L.}\ \bibnamefont
  {Goldstein}},\ }\bibfield  {title} {\bibinfo {title} {Measurement of the
  rugged invariants of magnetohydrodynamic turbulence in the solar wind},\
  }\href {https://doi.org/10.1029/JA087iA08p06011} {\bibfield  {journal}
  {\bibinfo  {journal} {\jgr}\ }\textbf {\bibinfo {volume} {87}},\ \bibinfo
  {pages} {6011} (\bibinfo {year} {1982})}\BibitemShut {NoStop}%
\bibitem [{\citenamefont {Batchelor}(1953)}]{batchelor1953theory}%
  \BibitemOpen
  \bibfield  {author} {\bibinfo {author} {\bibfnamefont {G.~K.}\ \bibnamefont
  {Batchelor}},\ }\href@noop {} {\emph {\bibinfo {title} {The theory of
  homogeneous turbulence}}}\ (\bibinfo  {publisher} {Cambridge university
  press},\ \bibinfo {year} {1953})\BibitemShut {NoStop}%
\bibitem [{\citenamefont {Sorriso-Valvo}\ \emph {et~al.}(2018)\citenamefont
  {Sorriso-Valvo}, \citenamefont {Carbone}, \citenamefont {Perri},
  \citenamefont {Greco}, \citenamefont {Marino},\ and\ \citenamefont
  {Bruno}}]{sorrisovalvo2018statistical}%
  \BibitemOpen
  \bibfield  {author} {\bibinfo {author} {\bibfnamefont {L.}~\bibnamefont
  {Sorriso-Valvo}}, \bibinfo {author} {\bibfnamefont {F.}~\bibnamefont
  {Carbone}}, \bibinfo {author} {\bibfnamefont {S.}~\bibnamefont {Perri}},
  \bibinfo {author} {\bibfnamefont {A.}~\bibnamefont {Greco}}, \bibinfo
  {author} {\bibfnamefont {R.}~\bibnamefont {Marino}},\ and\ \bibinfo {author}
  {\bibfnamefont {R.}~\bibnamefont {Bruno}},\ }\bibfield  {title} {\bibinfo
  {title} {On the statistical properties of turbulent energy transfer rate in
  the inner heliosphere},\ }\href {https://doi.org/10.1007/s11207-017-1229-6}
  {\bibfield  {journal} {\bibinfo  {journal} {Solar Physics}\ }\textbf
  {\bibinfo {volume} {293}},\ \bibinfo {pages} {10} (\bibinfo {year}
  {2018})}\BibitemShut {NoStop}%
\bibitem [{\citenamefont {Bandyopadhyay}\ \emph {et~al.}(2020)\citenamefont
  {Bandyopadhyay}, \citenamefont {Sorriso-Valvo}, \citenamefont {Chasapis},
  \citenamefont {Hellinger}, \citenamefont {Matthaeus}, \citenamefont
  {Verdini}, \citenamefont {Landi}, \citenamefont {Franci}, \citenamefont
  {Matteini}, \citenamefont {Giles}, \citenamefont {Gershman}, \citenamefont
  {Moore}, \citenamefont {Pollock}, \citenamefont {Russell}, \citenamefont
  {Strangeway}, \citenamefont {Torbert},\ and\ \citenamefont
  {Burch}}]{bandyopadhyay2020situ}%
  \BibitemOpen
  \bibfield  {author} {\bibinfo {author} {\bibfnamefont {R.}~\bibnamefont
  {Bandyopadhyay}}, \bibinfo {author} {\bibfnamefont {L.}~\bibnamefont
  {Sorriso-Valvo}}, \bibinfo {author} {\bibfnamefont {A.}~\bibnamefont
  {Chasapis}}, \bibinfo {author} {\bibfnamefont {P.}~\bibnamefont {Hellinger}},
  \bibinfo {author} {\bibfnamefont {W.~H.}\ \bibnamefont {Matthaeus}}, \bibinfo
  {author} {\bibfnamefont {A.}~\bibnamefont {Verdini}}, \bibinfo {author}
  {\bibfnamefont {S.}~\bibnamefont {Landi}}, \bibinfo {author} {\bibfnamefont
  {L.}~\bibnamefont {Franci}}, \bibinfo {author} {\bibfnamefont
  {L.}~\bibnamefont {Matteini}}, \bibinfo {author} {\bibfnamefont {B.~L.}\
  \bibnamefont {Giles}}, \bibinfo {author} {\bibfnamefont {D.~J.}\ \bibnamefont
  {Gershman}}, \bibinfo {author} {\bibfnamefont {T.~E.}\ \bibnamefont {Moore}},
  \bibinfo {author} {\bibfnamefont {C.~J.}\ \bibnamefont {Pollock}}, \bibinfo
  {author} {\bibfnamefont {C.~T.}\ \bibnamefont {Russell}}, \bibinfo {author}
  {\bibfnamefont {R.~J.}\ \bibnamefont {Strangeway}}, \bibinfo {author}
  {\bibfnamefont {R.~B.}\ \bibnamefont {Torbert}},\ and\ \bibinfo {author}
  {\bibfnamefont {J.~L.}\ \bibnamefont {Burch}},\ }\bibfield  {title} {\bibinfo
  {title} {In situ observation of hall magnetohydrodynamic cascade in space
  plasma},\ }\href {https://doi.org/10.1103/PhysRevLett.124.225101} {\bibfield
  {journal} {\bibinfo  {journal} {Phys. Rev. Lett.}\ }\textbf {\bibinfo
  {volume} {124}},\ \bibinfo {pages} {225101} (\bibinfo {year}
  {2020})}\BibitemShut {NoStop}%
\bibitem [{\citenamefont {Marino}\ and\ \citenamefont
  {Sorriso-Valvo}(2023)}]{MarinoSorriso23}%
  \BibitemOpen
  \bibfield  {author} {\bibinfo {author} {\bibfnamefont {R.}~\bibnamefont
  {Marino}}\ and\ \bibinfo {author} {\bibfnamefont {L.}~\bibnamefont
  {Sorriso-Valvo}},\ }\bibfield  {title} {\bibinfo {title} {Scaling laws for
  the energy transfer in space plasma turbulence},\ }\href
  {https://doi.org/https://doi.org/10.1016/j.physrep.2022.12.001} {\bibfield
  {journal} {\bibinfo  {journal} {Physics Reports}\ }\textbf {\bibinfo {volume}
  {1006}},\ \bibinfo {pages} {1} (\bibinfo {year} {2023})}\BibitemShut
  {NoStop}%
\bibitem [{\citenamefont {{Wang}}\ \emph {et~al.}(2022)\citenamefont {{Wang}},
  \citenamefont {{Chhiber}}, \citenamefont {{Adhikari}}, \citenamefont
  {{Yang}}, \citenamefont {{Bandyopadhyay}}, \citenamefont {{Shay}},
  \citenamefont {{Oughton}}, \citenamefont {{Matthaeus}},\ and\ \citenamefont
  {{Cuesta}}}]{WangEA22}%
  \BibitemOpen
  \bibfield  {author} {\bibinfo {author} {\bibfnamefont {Y.}~\bibnamefont
  {{Wang}}}, \bibinfo {author} {\bibfnamefont {R.}~\bibnamefont {{Chhiber}}},
  \bibinfo {author} {\bibfnamefont {S.}~\bibnamefont {{Adhikari}}}, \bibinfo
  {author} {\bibfnamefont {Y.}~\bibnamefont {{Yang}}}, \bibinfo {author}
  {\bibfnamefont {R.}~\bibnamefont {{Bandyopadhyay}}}, \bibinfo {author}
  {\bibfnamefont {M.~A.}\ \bibnamefont {{Shay}}}, \bibinfo {author}
  {\bibfnamefont {S.}~\bibnamefont {{Oughton}}}, \bibinfo {author}
  {\bibfnamefont {W.~H.}\ \bibnamefont {{Matthaeus}}},\ and\ \bibinfo {author}
  {\bibfnamefont {M.~E.}\ \bibnamefont {{Cuesta}}},\ }\bibfield  {title}
  {\bibinfo {title} {{Strategies for Determining the Cascade Rate in MHD
  Turbulence: Isotropy, Anisotropy, and Spacecraft Sampling}},\ }\href
  {https://doi.org/10.3847/1538-4357/ac8f90} {\bibfield  {journal} {\bibinfo
  {journal} {Astrophys. J.}\ }\textbf {\bibinfo {volume} {937}},\ \bibinfo
  {eid} {76} (\bibinfo {year} {2022})},\ \Eprint
  {https://arxiv.org/abs/2209.00208} {arXiv:2209.00208 [physics.space-ph]}
  \BibitemShut {NoStop}%
\bibitem [{\citenamefont {Politano}\ \emph {et~al.}(1998)\citenamefont
  {Politano}, \citenamefont {Pouquet},\ and\ \citenamefont
  {Carbone}}]{PolitanoEA98}%
  \BibitemOpen
  \bibfield  {author} {\bibinfo {author} {\bibfnamefont {H.}~\bibnamefont
  {Politano}}, \bibinfo {author} {\bibfnamefont {A.}~\bibnamefont {Pouquet}},\
  and\ \bibinfo {author} {\bibfnamefont {V.}~\bibnamefont {Carbone}},\
  }\bibfield  {title} {\bibinfo {title} {Determination of anomalous exponents
  of structure functions in two-dimensional magnetohydrodynamic turbulence},\
  }\href {https://doi.org/10.1209/epl/i1998-00391-2} {\bibfield  {journal}
  {\bibinfo  {journal} {Europhys. Lett.}\ }\textbf {\bibinfo {volume} {43}},\
  \bibinfo {pages} {516} (\bibinfo {year} {1998})}\BibitemShut {NoStop}%
\bibitem [{\citenamefont {Osman}\ \emph {et~al.}(2011)\citenamefont {Osman},
  \citenamefont {Wan}, \citenamefont {Matthaeus}, \citenamefont {Weygand},\
  and\ \citenamefont {Dasso}}]{OsmanEA11-3rd}%
  \BibitemOpen
  \bibfield  {author} {\bibinfo {author} {\bibfnamefont {K.~T.}\ \bibnamefont
  {Osman}}, \bibinfo {author} {\bibfnamefont {M.}~\bibnamefont {Wan}}, \bibinfo
  {author} {\bibfnamefont {W.~H.}\ \bibnamefont {Matthaeus}}, \bibinfo {author}
  {\bibfnamefont {J.~M.}\ \bibnamefont {Weygand}},\ and\ \bibinfo {author}
  {\bibfnamefont {S.}~\bibnamefont {Dasso}},\ }\bibfield  {title} {\bibinfo
  {title} {Anisotropic third-moment estimates of the energy cascade in solar
  wind turbulence using multispacecraft data},\ }\href
  {https://doi.org/10.1103/PhysRevLett.107.165001} {\bibfield  {journal}
  {\bibinfo  {journal} {Phys.\ Rev.\ Lett.}\ }\textbf {\bibinfo {volume}
  {107}},\ \bibinfo {pages} {165001} (\bibinfo {year} {2011})}\BibitemShut
  {NoStop}%
\bibitem [{\citenamefont {{Bandyopadhyay}}\ \emph {et~al.}(2020)\citenamefont
  {{Bandyopadhyay}}, \citenamefont {{Goldstein}}, \citenamefont {{Maruca}},
  \citenamefont {{Matthaeus}}, \citenamefont {{Parashar}}, \citenamefont
  {{Ruffolo}}, \citenamefont {{Chhiber}}, \citenamefont {{Usmanov}},
  \citenamefont {{Chasapis}}, \citenamefont {{Qudsi}}, \citenamefont {{Bale}},
  \citenamefont {{Bonnell}}, \citenamefont {{Dudok de Wit}}, \citenamefont
  {{Goetz}}, \citenamefont {{Harvey}}, \citenamefont {{MacDowall}},
  \citenamefont {{Malaspina}}, \citenamefont {{Pulupa}}, \citenamefont
  {{Kasper}}, \citenamefont {{Korreck}}, \citenamefont {{Case}}, \citenamefont
  {{Stevens}}, \citenamefont {{Whittlesey}}, \citenamefont {{Larson}},
  \citenamefont {{Livi}}, \citenamefont {{Klein}}, \citenamefont {{Velli}},\
  and\ \citenamefont {{Raouafi}}}]{BandyopadhyayEA20-PSP3rd}%
  \BibitemOpen
  \bibfield  {author} {\bibinfo {author} {\bibfnamefont {R.}~\bibnamefont
  {{Bandyopadhyay}}}, \bibinfo {author} {\bibfnamefont {M.~L.}\ \bibnamefont
  {{Goldstein}}}, \bibinfo {author} {\bibfnamefont {B.~A.}\ \bibnamefont
  {{Maruca}}}, \bibinfo {author} {\bibfnamefont {W.~H.}\ \bibnamefont
  {{Matthaeus}}}, \bibinfo {author} {\bibfnamefont {T.~N.}\ \bibnamefont
  {{Parashar}}}, \bibinfo {author} {\bibfnamefont {D.}~\bibnamefont
  {{Ruffolo}}}, \bibinfo {author} {\bibfnamefont {R.}~\bibnamefont
  {{Chhiber}}}, \bibinfo {author} {\bibfnamefont {A.}~\bibnamefont
  {{Usmanov}}}, \bibinfo {author} {\bibfnamefont {A.}~\bibnamefont
  {{Chasapis}}}, \bibinfo {author} {\bibfnamefont {R.}~\bibnamefont {{Qudsi}}},
  \bibinfo {author} {\bibfnamefont {S.~D.}\ \bibnamefont {{Bale}}}, \bibinfo
  {author} {\bibfnamefont {J.~W.}\ \bibnamefont {{Bonnell}}}, \bibinfo {author}
  {\bibfnamefont {T.}~\bibnamefont {{Dudok de Wit}}}, \bibinfo {author}
  {\bibfnamefont {K.}~\bibnamefont {{Goetz}}}, \bibinfo {author} {\bibfnamefont
  {P.~R.}\ \bibnamefont {{Harvey}}}, \bibinfo {author} {\bibfnamefont {R.~J.}\
  \bibnamefont {{MacDowall}}}, \bibinfo {author} {\bibfnamefont {D.~M.}\
  \bibnamefont {{Malaspina}}}, \bibinfo {author} {\bibfnamefont
  {M.}~\bibnamefont {{Pulupa}}}, \bibinfo {author} {\bibfnamefont {J.~C.}\
  \bibnamefont {{Kasper}}}, \bibinfo {author} {\bibfnamefont {K.~E.}\
  \bibnamefont {{Korreck}}}, \bibinfo {author} {\bibfnamefont {A.~W.}\
  \bibnamefont {{Case}}}, \bibinfo {author} {\bibfnamefont {M.}~\bibnamefont
  {{Stevens}}}, \bibinfo {author} {\bibfnamefont {P.}~\bibnamefont
  {{Whittlesey}}}, \bibinfo {author} {\bibfnamefont {D.}~\bibnamefont
  {{Larson}}}, \bibinfo {author} {\bibfnamefont {R.}~\bibnamefont {{Livi}}},
  \bibinfo {author} {\bibfnamefont {K.~G.}\ \bibnamefont {{Klein}}}, \bibinfo
  {author} {\bibfnamefont {M.}~\bibnamefont {{Velli}}},\ and\ \bibinfo {author}
  {\bibfnamefont {N.}~\bibnamefont {{Raouafi}}},\ }\bibfield  {title} {\bibinfo
  {title} {{Enhanced Energy Transfer Rate in Solar Wind Turbulence Observed
  near the Sun from Parker Solar Probe}},\ }\href
  {https://doi.org/10.3847/1538-4365/ab5dae} {\bibfield  {journal} {\bibinfo
  {journal} {Astrophys. J. Suppl.}\ }\textbf {\bibinfo {volume} {246}},\
  \bibinfo {eid} {48} (\bibinfo {year} {2020})},\ \Eprint
  {https://arxiv.org/abs/1912.02959} {arXiv:1912.02959 [physics.space-ph]}
  \BibitemShut {NoStop}%
\bibitem [{\citenamefont {Dunlop}\ \emph {et~al.}(2002)\citenamefont {Dunlop},
  \citenamefont {Balogh}, \citenamefont {Glassmeier},\ and\ \citenamefont
  {Robert}}]{dunlop2002four}%
  \BibitemOpen
  \bibfield  {author} {\bibinfo {author} {\bibfnamefont {M.}~\bibnamefont
  {Dunlop}}, \bibinfo {author} {\bibfnamefont {A.}~\bibnamefont {Balogh}},
  \bibinfo {author} {\bibfnamefont {K.-H.}\ \bibnamefont {Glassmeier}},\ and\
  \bibinfo {author} {\bibfnamefont {P.}~\bibnamefont {Robert}},\ }\bibfield
  {title} {\bibinfo {title} {Four-point cluster application of magnetic field
  analysis tools: The curlometer},\ }\href@noop {} {\bibfield  {journal}
  {\bibinfo  {journal} {Journal of Geophysical Research: Space Physics}\
  }\textbf {\bibinfo {volume} {107}},\ \bibinfo {pages} {SMP} (\bibinfo {year}
  {2002})}\BibitemShut {NoStop}%
\bibitem [{\citenamefont {Verdini}\ \emph {et~al.}(2015)\citenamefont
  {Verdini}, \citenamefont {Grappin}, \citenamefont {Hellinger}, \citenamefont
  {Landi},\ and\ \citenamefont {Müller}}]{verdini2015anisotropy}%
  \BibitemOpen
  \bibfield  {author} {\bibinfo {author} {\bibfnamefont {A.}~\bibnamefont
  {Verdini}}, \bibinfo {author} {\bibfnamefont {R.}~\bibnamefont {Grappin}},
  \bibinfo {author} {\bibfnamefont {P.}~\bibnamefont {Hellinger}}, \bibinfo
  {author} {\bibfnamefont {S.}~\bibnamefont {Landi}},\ and\ \bibinfo {author}
  {\bibfnamefont {W.~C.}\ \bibnamefont {Müller}},\ }\bibfield  {title}
  {\bibinfo {title} {Anisotropy of third-order structure functions in mhd
  turbulence},\ }\href {https://doi.org/10.1088/0004-637X/804/2/119} {\bibfield
   {journal} {\bibinfo  {journal} {The Astrophysical Journal}\ }\textbf
  {\bibinfo {volume} {804}},\ \bibinfo {pages} {119} (\bibinfo {year}
  {2015})}\BibitemShut {NoStop}%
\bibitem [{\citenamefont {Bandyopadhyay}\ \emph {et~al.}(2018)\citenamefont
  {Bandyopadhyay}, \citenamefont {Chasapis}, \citenamefont {Chhiber},
  \citenamefont {Parashar}, \citenamefont {Matthaeus}, \citenamefont {Shay},
  \citenamefont {Maruca}, \citenamefont {Burch}, \citenamefont {Moore},
  \citenamefont {Pollock}, \citenamefont {Giles}, \citenamefont {Paterson},
  \citenamefont {Dorelli}, \citenamefont {Gershman}, \citenamefont {Torbert},
  \citenamefont {Russell},\ and\ \citenamefont
  {Strangeway}}]{bandyopadhyay2018incompressive}%
  \BibitemOpen
  \bibfield  {author} {\bibinfo {author} {\bibfnamefont {R.}~\bibnamefont
  {Bandyopadhyay}}, \bibinfo {author} {\bibfnamefont {A.}~\bibnamefont
  {Chasapis}}, \bibinfo {author} {\bibfnamefont {R.}~\bibnamefont {Chhiber}},
  \bibinfo {author} {\bibfnamefont {T.~N.}\ \bibnamefont {Parashar}}, \bibinfo
  {author} {\bibfnamefont {W.~H.}\ \bibnamefont {Matthaeus}}, \bibinfo {author}
  {\bibfnamefont {M.~A.}\ \bibnamefont {Shay}}, \bibinfo {author}
  {\bibfnamefont {B.~A.}\ \bibnamefont {Maruca}}, \bibinfo {author}
  {\bibfnamefont {J.~L.}\ \bibnamefont {Burch}}, \bibinfo {author}
  {\bibfnamefont {T.~E.}\ \bibnamefont {Moore}}, \bibinfo {author}
  {\bibfnamefont {C.~J.}\ \bibnamefont {Pollock}}, \bibinfo {author}
  {\bibfnamefont {B.~L.}\ \bibnamefont {Giles}}, \bibinfo {author}
  {\bibfnamefont {W.~R.}\ \bibnamefont {Paterson}}, \bibinfo {author}
  {\bibfnamefont {J.}~\bibnamefont {Dorelli}}, \bibinfo {author} {\bibfnamefont
  {D.~J.}\ \bibnamefont {Gershman}}, \bibinfo {author} {\bibfnamefont {R.~B.}\
  \bibnamefont {Torbert}}, \bibinfo {author} {\bibfnamefont {C.~T.}\
  \bibnamefont {Russell}},\ and\ \bibinfo {author} {\bibfnamefont {R.~J.}\
  \bibnamefont {Strangeway}},\ }\bibfield  {title} {\bibinfo {title}
  {Incompressive energy transfer in the earth’s magnetosheath: Magnetospheric
  multiscale observations},\ }\href {https://doi.org/10.3847/1538-4357/aade04}
  {\bibfield  {journal} {\bibinfo  {journal} {The Astrophysical Journal}\
  }\textbf {\bibinfo {volume} {866}},\ \bibinfo {pages} {106} (\bibinfo {year}
  {2018})}\BibitemShut {NoStop}%
\bibitem [{\citenamefont {Maruca}\ \emph {et~al.}(2021)\citenamefont {Maruca},
  \citenamefont {Agudelo~Rueda}, \citenamefont {Bandyopadhyay}, \citenamefont
  {Bianco}, \citenamefont {Chasapis}, \citenamefont {Chhiber}, \citenamefont
  {DeWeese}, \citenamefont {Matthaeus}, \citenamefont {Miles}, \citenamefont
  {Qudsi}, \citenamefont {Richardson}, \citenamefont {Servidio}, \citenamefont
  {Shay}, \citenamefont {Sundkvist}, \citenamefont {Verscharen}, \citenamefont
  {Vines}, \citenamefont {Westlake},\ and\ \citenamefont
  {Wicks}}]{maruca2021magnetore}%
  \BibitemOpen
  \bibfield  {author} {\bibinfo {author} {\bibfnamefont {B.~A.}\ \bibnamefont
  {Maruca}}, \bibinfo {author} {\bibfnamefont {J.~A.}\ \bibnamefont
  {Agudelo~Rueda}}, \bibinfo {author} {\bibfnamefont {R.}~\bibnamefont
  {Bandyopadhyay}}, \bibinfo {author} {\bibfnamefont {F.~B.}\ \bibnamefont
  {Bianco}}, \bibinfo {author} {\bibfnamefont {A.}~\bibnamefont {Chasapis}},
  \bibinfo {author} {\bibfnamefont {R.}~\bibnamefont {Chhiber}}, \bibinfo
  {author} {\bibfnamefont {H.}~\bibnamefont {DeWeese}}, \bibinfo {author}
  {\bibfnamefont {W.~H.}\ \bibnamefont {Matthaeus}}, \bibinfo {author}
  {\bibfnamefont {D.~M.}\ \bibnamefont {Miles}}, \bibinfo {author}
  {\bibfnamefont {R.~A.}\ \bibnamefont {Qudsi}}, \bibinfo {author}
  {\bibfnamefont {M.~J.}\ \bibnamefont {Richardson}}, \bibinfo {author}
  {\bibfnamefont {S.}~\bibnamefont {Servidio}}, \bibinfo {author}
  {\bibfnamefont {M.~A.}\ \bibnamefont {Shay}}, \bibinfo {author}
  {\bibfnamefont {D.}~\bibnamefont {Sundkvist}}, \bibinfo {author}
  {\bibfnamefont {D.}~\bibnamefont {Verscharen}}, \bibinfo {author}
  {\bibfnamefont {S.~K.}\ \bibnamefont {Vines}}, \bibinfo {author}
  {\bibfnamefont {J.~H.}\ \bibnamefont {Westlake}},\ and\ \bibinfo {author}
  {\bibfnamefont {R.~T.}\ \bibnamefont {Wicks}},\ }\bibfield  {title} {\bibinfo
  {title} {Magnetore: Mapping the 3-d magnetic structure of the solar wind
  using a large constellation of nanosatellites},\ }\bibfield  {journal}
  {\bibinfo  {journal} {Frontiers in Astronomy and Space Sciences}\ }\textbf
  {\bibinfo {volume} {8}},\ \href {https://doi.org/10.3389/fspas.2021.665885}
  {10.3389/fspas.2021.665885} (\bibinfo {year} {2021})\BibitemShut {NoStop}%
\end{thebibliography}%
\end{document}